# DCE-Qnet: Deep Network Quantification of Dynamic Contrast Enhanced (DCE) MRI

Ouri Cohen*[1,†], Soudabeh Kargar[1,†], Sungmin Woo[2], Alberto Vargas[2], Ricardo Otazo[1,2]

[1]Department of Medical Physics, Memorial Sloan Kettering Cancer Center, New York, NY, USA

[2]Department of Radiology, Memorial Sloan Kettering Cancer Center, New York, NY, USA

**Correspondence to:** Ouri Cohen, coheno1@mskcc.org, Memorial Sloan Kettering Cancer Center, 320 East 61st St, New York, NY, 10025, USA.

[†]Equal contributors

**Word Count:** 5723

**Research Article**

# ABSTRACT

**Introduction:** Quantification of dynamic contrast-enhanced (DCE)-MRI has the potential to provide valuable clinical information, but robust pharmacokinetic modeling remains a challenge for clinical adoption.

**Methods:** A 7-layer neural network called DCE-Qnet was trained on simulated DCE-MRI signals derived from the Extended Tofts model with the Parker arterial input function. Network training incorporated B1 inhomogeneities to estimate perfusion ($K^{trans}$, $v_p$, $v_e$), tissue T1 relaxation, proton density and bolus arrival time (BAT). The accuracy was tested in a digital phantom in comparison to a conventional nonlinear least-squares fitting (NLSQ). In vivo testing was conducted in 10 healthy subjects. Regions of interest in the cervix and uterine myometrium were used to calculate the inter-subject variability. The clinical utility was demonstrated on a cervical cancer patient. Test-retest experiments were used to assess reproducibility of the parameter maps in the tumor.

**Results:** The DCE-Qnet reconstruction outperformed NLSQ in the phantom. The coefficient of variation (CV) in the healthy cervix varied between 5-51% depending on the parameter. Parameter values in the tumor agreed with previous studies despite differences in methodology. The CV in the tumor varied between 1-47%.

**Conclusion:** The proposed approach provides comprehensive DCE-MRI quantification from a single acquisition. DCE-Qnet eliminates the need for separate T1 scan or BAT processing, leading to a reduction of 10 minutes per scan and more accurate quantification.

**Keywords**: DCE, deep learning, DRONE, neural network

# 1. Introduction

Quantitative parametric maps derived from dynamic contrast-enhanced (DCE)-MRI data can provide important information for cancer diagnosis and early treatment response evaluation. For example, quantitative DCE-MRI can be used to study tumor vascularity and identify tumors that are better perfused and oxygenated and thus more sensitive to chemotherapy or radiotherapy [1]. However, despite extensive research work and promising results, quantitative perfusion parameters still play a very limited role in routine clinical practice [2]. One of the main causes has been the challenging pharmacokinetic modeling and quantification involved in DCE analysis. The DCE perfusion parameters of interest are sensitive to patient-specific variables, such as T1, arterial input function (AIF), bolus arrival time (BAT), blood hematocrit and others, which must be known for accurate quantification. While some of these parameters can be estimated using separate scans or measurements, misregistration between the separate acquisitions (induced by motion, for example) can introduce errors into the maps. Similarly, most of the methods for BAT calculation rely on a subjective delineation of some threshold to determine the contrast agent (CA) onset [3], [4]. In short, the dependence of the signal on multiple parameters and the complexity of the pharmacokinetic models is a significant challenge for reproducible quantification.

Deep neural networks (DNN) are attractive for DCE-MRI quantification, as they can efficiently model complex, nonlinear functions [5] and are thus able to map measured signals to the underlying parameters even with limited training data [6], [7], [8]. This allows DNNs to efficiently solve the inverse problem [9], [10] and correctly estimate the underlying parameters. Recent work demonstrated the usefulness of DNNs for DCE-MRI estimation [4], [11], [12], [13]. However, most of these studies required a separate T1 acquisition and BAT analysis and the use of patient-derived AIF curves, which may not be universally available, to train the neural network.

This work presents a comprehensive DNN-based quantification method for DCE-MRI data called DCE-Qnet that estimates quantitative T1, proton density (PD), and BAT maps, in addition to perfusion parameters ($K^{trans}$, $v_p$, $v_e$), using a single acquisition, and therefore eliminates the need for a separate T1 acquisition or BAT analysis. DCE-Qnet also presents a novel way to map

T1 from contrast enhancement only. The methodology proposed in DCE-Qnet is inspired by the DRONE approach [6], which utilizes a DNN trained with simulated, physics-derived Bloch equations to perform T1 and T2 mapping. DCE-Qnet is evaluated using digital phantoms with realistic T1, BAT and perfusion parameter values, and compared to a standard fitting algorithm. The clinical utility of DCE-Qnet is demonstrated in 10 healthy subjects and one patient with cervical cancer.

## 2. Methods

### 2.1. Neural network architecture and training

The mathematical framework and notation of the extended Tofts model used in this work are described in the Appendix section. A 7-dimensional neural network was defined consisting of a 60-nodes input layer, five 300-node hidden layers and a 5-node output layer (Figure 1). A set of 100,000 tissue parameters were sampled from the following ranges: $K^{trans}$=[0, 0.5] 1/min, $v_p$=[0, 70]%, $v_e$=[0, 80]%, $T_{10}$=[0, 3000]ms, BAT=[25, 175] seconds , B1=[0, 2] as a scaling of the nominal B1. Sampling was accomplished using latin-hypercube (LH) sampling, a statistical method for generating random but relatively uniformly distributed samples in each dimension [14]. The transmitter $B_1$ field was included in network training to account for inevitable $B_1$ inhomogeneities but excluded from network output to avoid additional burden in the estimation process. For each set of parameters, a DCE-MRI experiment was simulated using Eqns. (A1)-(A5). The parameters of the Parker AIF were obtained from Ref. [15] and the resulting signals were normalized to have a $l_2$-norm of 1. The equilibrium magnetization ($S_0$) in the signal (Eqn. A5) is a scaling factor that is removed during normalization and was hence set to 1 in the analysis. The BAT value was defined as the time-shift which minimized the difference between the measured and fitted curves for each voxel. A validation set was defined using 20% of the generated data to quantify the quality of the network training with the remainder used for training. The DCE-MRI simulation was implemented on an Nvidia RTX2080 Ti graphics processing unit (GPU) (Nvidia Corp. Santa Clara, CA) with 11 GB of memory. The loss was defined as the mean absolute error and the network was trained with the Adam optimizer [16] with a batch size of 1000 and an adaptive learning rate with weight decay of $10^{-4}$. Zero mean Gaussian noise was added during training to promote robust learning [17] and the network was trained for 3000 epochs.

The estimated parameters were used in combination with the measured data to quantify the proton density (PD). As in MR fingerprinting (MRF), the PD of each voxel was calculated as a scaling of the reconstructed data using the expression [18]: $PD = \frac{m^T d}{\|d\|^2}$ where **m** is the measured signal and **d** is the signal synthesized from the tissue parameter values found by the network.

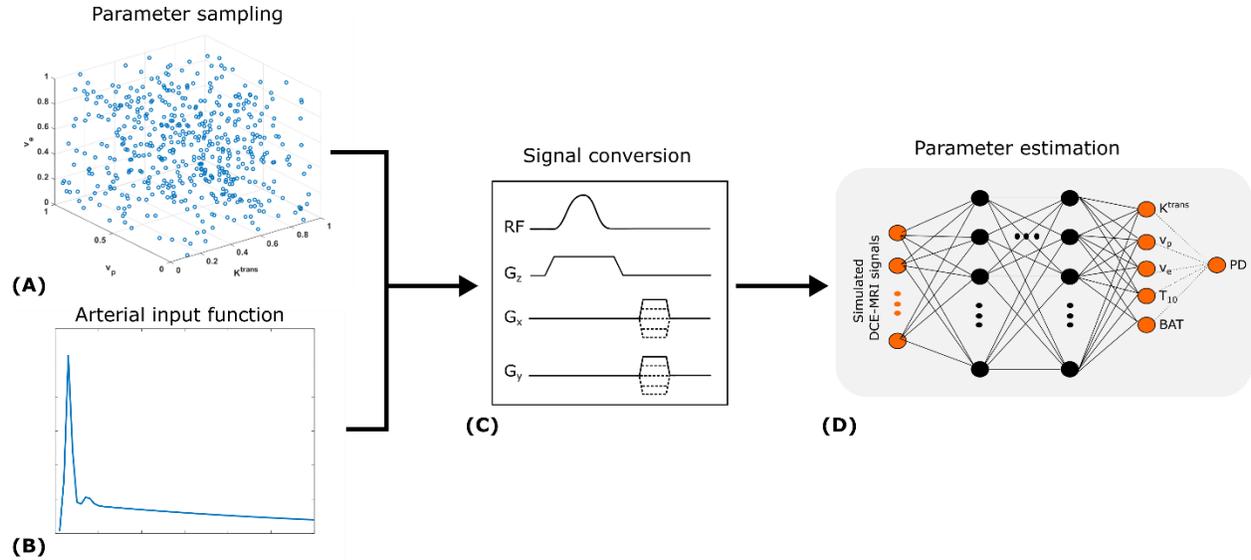

Figure 1: Main components of the DCE-Qnet estimation process. (A) Sampling of the DCE parameters shown for 3 out of 6 parameters for visualization purposes. The parameter values were generated by LH sampling of the 6-dimensional space. (B) The population-based Parker AIF was used to generate the concentration-time curves for each of the DCE parameter combinations. (C) Concentration-time curves were converted to signal intensity values using the signal model defined in Eqn. (A5). (D) Diagram of the neural network used. Signal intensity curves generated from the sampled DCE parameters were used as input to the network during training. The network learned to map the signal intensity curves to the underlying tissue parameters. The proton density was then calculated from measured and synthesized signals obtained with the estimated DCE parameters.

## 2.2. Numerical simulations

### 2.2.1. Reconstruction accuracy in comparison to alternative reconstruction method

The accuracy of the DNN reconstruction was tested in a custom digital phantom consisting of 2500 pixels, each corresponding to a set of DCE parameters selected at random from a uniform distribution. A DCE-MRI acquisition with N=60 time-steps was simulated, as described in the Appendix section, for each pixel and the resulting signal used as input to the trained DNN. The simulations used the same scan protocol as that of the healthy subject (see section 2.3.1). The Pearson correlation between the DNN-estimated parameters and the reference values was calculated, along with the RMSE. The linear-least-squares best-fit line between the reference and estimated parameters was also calculated to evaluate the deviation from linearity.

The accuracy of the DNN reconstruction in the phantom was compared to a standard reconstruction using nonlinear-least-squares (NLSQ) fitting implemented using the lsqcurvefit() function (Matlab R2018b, The Mathworks, Natick, MA). The algorithm was initialized with the following values: $K^{trans}$=0.2, $v_p$=40%, $v_e$=40%, T1=1000 ms, BAT=50. The estimated values for each parameter were compared to the true values. To simulate the effect of a separate B1 scan, T1 scan and BAT analysis, as in conventional DCE processing [19], in a second experiment, the fitting algorithm was also provided with the true B1, T1 and BAT values with the fitting accuracy quantified for the remaining parameters. Finally, the NLSQ reconstruction was compared to a DNN reconstruction. All parameters were simultaneously estimated in the DNN reconstruction. Given the increased difficulty in fitting high dimensional signals and to ensure a fair comparison between the NLSQ and DNN reconstructions, the true B1, T1 and BAT values were only used to initialize the NLSQ algorithm rather than reduce the degrees-of-freedom.

### 2.2.2. Effect of injected training noise

There is a tradeoff between the noise injected during training and the robustness of the DNN reconstruction to measurement noise. Noise injected in training serves to focus the network on actual signal features rather than artifacts or noise, reducing the risk of overfitting and leading to smoother and more realistic parameter maps. On the other hand, excessive training noise may overwhelm small differences between signals arising from different parameters, thereby reducing the sensitivity of the reconstruction to small variations in the parameters. To quantify this tradeoff, three networks were trained with varying levels of injected training noise corresponding to a training SNR of approximately 1, 21 and 41 dB calculated as $SNR = 20 \cdot \log_{10}(S/N)$ with the signal $S$ obtained using the maximum value of the signals comprising the training dataset. The use of the maximum signal value provides a 'worse-case' analysis which elucidates the limits of the proposed approach. To test the robustness of each network to measurement noise, the phantom's simulated DCE signals were corrupted by zero mean Gaussian noise which was calibrated to obtain SNRs ranging from 20 to 80 dB. Each of the trained networks was used to reconstruct the phantom signals and the normalized RMS error (NRMSE) quantified for each parameter.

### 2.2.3. Effect of acquisition parameters on network quantification

Accurate modeling in quantitative MRI require knowledge of the scan parameters. To quantify the error induced by training the network with the wrong acquisition protocol, numerical simulations were conducted. A network was trained using the protocol used in the volunteers (described in section 3.3) and then used to reconstruct two datasets. The first dataset was generated with the same protocol as that used in the network training whereas the second dataset was generated with a different protocol, namely the protocol used in imaging the tumor patient.

## 2.3.    In vivo studies

### 2.3.1. MRI

In vivo acquisitions were performed on a 3T scanner (Premier, GE Healthcare, Milwaukee, WI) using a fat-suppressed T1-weighted stack-of-stars k-space acquisition (radial $k_y$-$k_x$ and Cartesian $k_z$ sampling) using the built-in transmit body coil and a 90-channel body receiver AIR coil. Acquired data were reconstructed using Golden-angle Radial Sparse Parallel (GRASP) [20], [21] method with a temporal resolution of 5 seconds.

Healthy subjects were scanned with the DCE-MRI acquisition with the following parameters: FOV= 260mm × 260mm, TR/TE= 3.4/1.6 msec, FA= 15°, scan time = 6.3 min, parallel imaging acceleration factor along the kz dimension = 2. A patient with cervical cancer was scanned using the following parameters, defined in accordance with our institution's clinical protocols which differed from the research protocol used in the healthy subject scan thereby necessitating a separately trained quantification network: FOV= 400mm×400mm, TR/TE= 3.2/1.5 msec, FA= 12°, scan time = 6.3 min.

The pre-contrast T1 map was obtained with a VFA sequence with flip angles=5°, 10°, 15°, 20° and the vendor's Bloch-Siegert mapping [22] used for B1 mapping with flip angle=10°, TE=2.4 ms and TR=164 ms.

### 2.3.2. Human subjects

Ten healthy female volunteers (ages 24-40, mean age 29±4.9 years) and a 41 years-old patient with cervical cancer were recruited for this study and provided IRB informed consent. Each subject was injected with 0.1 mmol/kg of Gadavist followed by a 20 ml saline flush at 3 ml/s using a mechanical power injector.

### 2.3.3. Data processing

A trained radiologist (S.W.) segmented the images into regions-of-interest (ROI) consisting of the cervix and outer uterine myometrium. In the cervical cancer subject, a tumor ROI was also defined. To validate the DNN-estimated maps, three representative voxels were selected in each dataset and the measured concentration-time curves were compared to concentration-time curves synthesized from the DNN-estimated parameters. The Pearson correlation was used to quantify the agreement between the measured and synthetized curves.

#### *2.3.3.1. Inter-subject variability*

The variability of the proposed method was quantified in the 10 healthy volunteers. The mean (μ) and SD (σ) for each ROI and tissue parameter were calculated across all subjects and the coefficient of variation, defined as CV=100×σ/μ, computed for each parameter and tissue type (cervix and myometrium).

#### *2.3.3.2. Comparison with NLSQ with measured B1 and T1*

The tissue maps obtained with DCE-Qnet in a healthy subject were compared to those obtained by NLSQ fitting using measured B1 and T1 maps and calculated BAT maps. The BAT was estimated by selecting the time point exhibiting the largest gradient in the signal. These parameters served as input to the NLSQ fitting algorithm to obtain the estimated $K^{trans}$, $v_p$ and $v_e$ maps.

### 2.3.4. Tumor patient

#### *2.3.4.1. Intra-subject reproducibility*

The in vivo reproducibility of the DNN quantification was assessed by test-retest experiments. The patient was scanned once with the DCE-MRI pulse sequence then scanned again 24 hours later. The data from both scans was reconstructed with the trained DNN and the CV used to calculate the agreement between the scans for each ROI.

# 3. Results

## 3.1. Numerical simulations

### 3.1.1. Quantification accuracy in comparison to alternative methods

The DCE-Qnet estimated parameters are shown in comparison to the reference values for each parameter in Figure 2A assuming a fixed B1, corresponding to estimation with a measured B1 map. The estimated and reference values were strongly correlated for all parameters ($R^2 \geq 0.85$). The best-fit line showed a maximum deviation from linearity of less than 6% across all parameters and little bias. The DCE-Qnet estimation of all parameters (Figure 2B) demonstrated increased variance although estimated parameters were still well correlated with the reference values. The NLSQ reconstruction performed poorly for simultaneous estimation of the 6 parameters, resulting in estimates that were uncorrelated to the reference values (Figure 2C). Similarly, NLSQ estimation of 4 parameters was generally poor (Figure 2D) even with a fixed B1. When the algorithm was initialized with the true $B_1$, BAT and T1 values the NLSQ estimation was markedly improved (Figure 2E), with the best-fit line showing a maximum deviation of less than 6%, similar to that of DCE-Qnet estimation of 5 parameters.

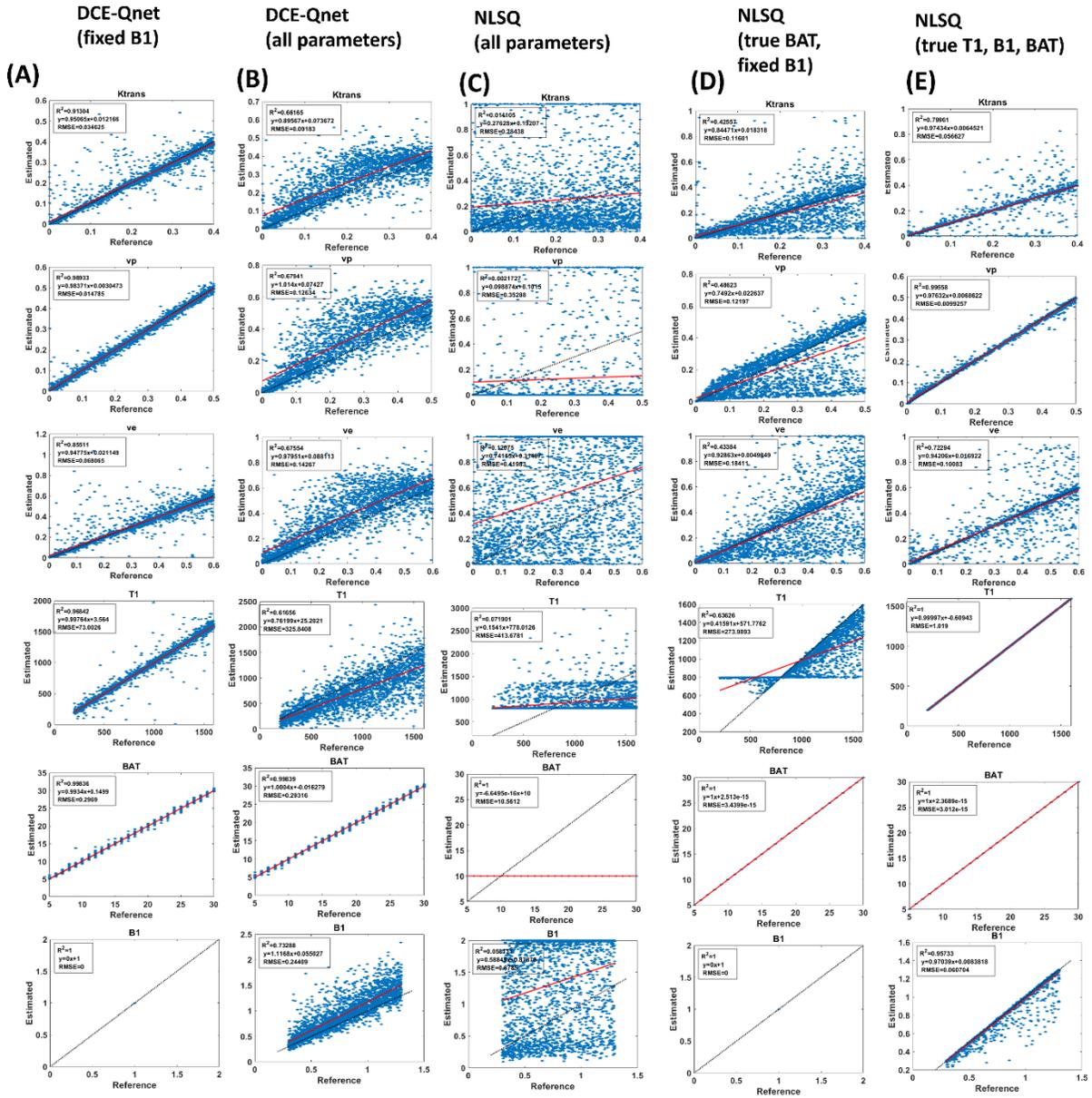

Figure 2: Reconstruction accuracy and comparison between NLSQ and DCE-Qnet. (A) Estimated vs reference comparison for parameters estimated with DCE-Qnet for a known fixed B1 equivalent to using a measured B1 map. (B) Estimated vs reference comparison for simultaneous DCE-Qnet estimation of all parameters. Although the variance was larger, the DCE-Qnet values were nevertheless strongly correlated with the reference values despite simultaneous estimation of all 6 parameters. (C) Estimated vs reference comparison for joint estimation of all parameters using the NLSQ algorithm. Compare with the DCE-

Qnet estimation shown in (B). (D) NLSQ estimation for fixed B1. (E) NLSQ estimation when the true B1, true T1 and true BAT were provided as inputs to the function.

### 3.1.2. Effect of injected training noise

The NRMSE of the parameter estimation obtained with DCE-Qnet networks trained with different noise levels is shown in Figure 3. Except for the BAT, the error of all parameters was predictably large at low SNR but dropped quickly with increasing SNR and converged to a level which varied depending on the injected noise level (Figure 3A). For example, the error converged at 20 dB for 1 dB training SNR but at 60 dB for 41 dB training SNR. Notably, greater injected noise resulted in faster convergence of the error albeit to a higher absolute error in comparison to a smaller noise injection.

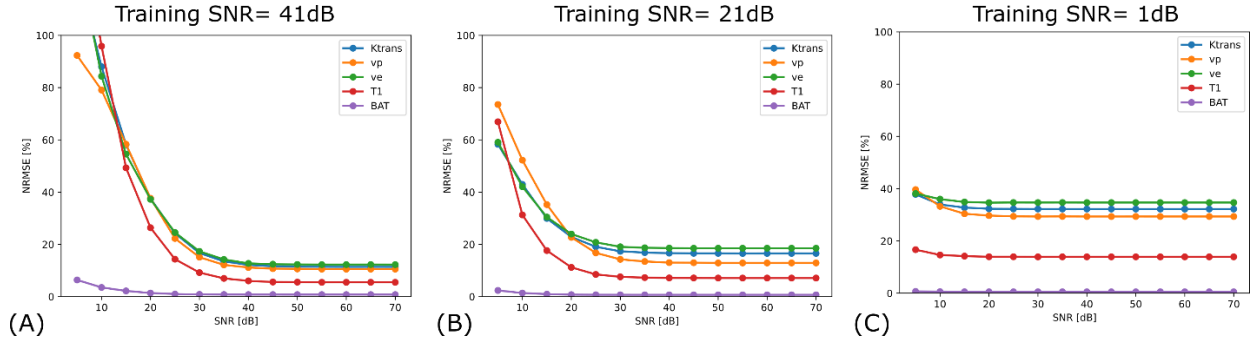

Figure 3: Normalized RMS error (NMRE) for different parameters as a function of SNR for networks trained with different injected noise levels. Shown is the output from the network trained with training SNR of (A) 41dB, (B) 21 dB and (C) 1dB. Note that higher levels of injected noise reduced the SNR at which the error converged albeit at the cost of a larger error.

### 3.1.3. Effect of acquisition parameters on network quantification

The tissue parameter values estimated with the network trained with the correct and wrong parameters in comparison to the reference values are shown in Supporting Figure S1. Using the incorrect protocol in the network training increased the reconstruction error validating the necessity of a separately trained network for acquisitions with different parameters.

### 3.2. In vivo subjects

### 3.2.1. Healthy subjects

Quantitative maps from a representative subject are shown in Figure 4A. The measured concentration-time curves (blue circles) for the three selected voxels are shown in Figure 4B along with the synthesized concentration-time curves in red. There was strong correlation (r ≥ 0.99) between the measured and synthesized curves. The distribution of values in the delineated ROIs for a representative subject is shown for each parameter in Figure 5. The distributions of the estimated parameters reflected differences in the tissue types, as expected. The distribution of all estimated parameters in the cervix for all subjects is shown in Figure 6. The mean±SD values across all subjects, along with the inter-subject CVs are listed in Table 1. The CV varied by parameter, ranging between 5 and 51% in the cervix, and between 8 and 88% in the uterine myometrium. The tissue maps obtained with DCE-Qnet in comparison to those obtained with NLSQ in a healthy subject are shown in Supporting Figure S2. The DCE-Qnet maps showed improved image quality and better delineation of the anatomy than NLSQ, despite the use of measured B1, T1 and calculated BAT maps in the NLSQ fitting.

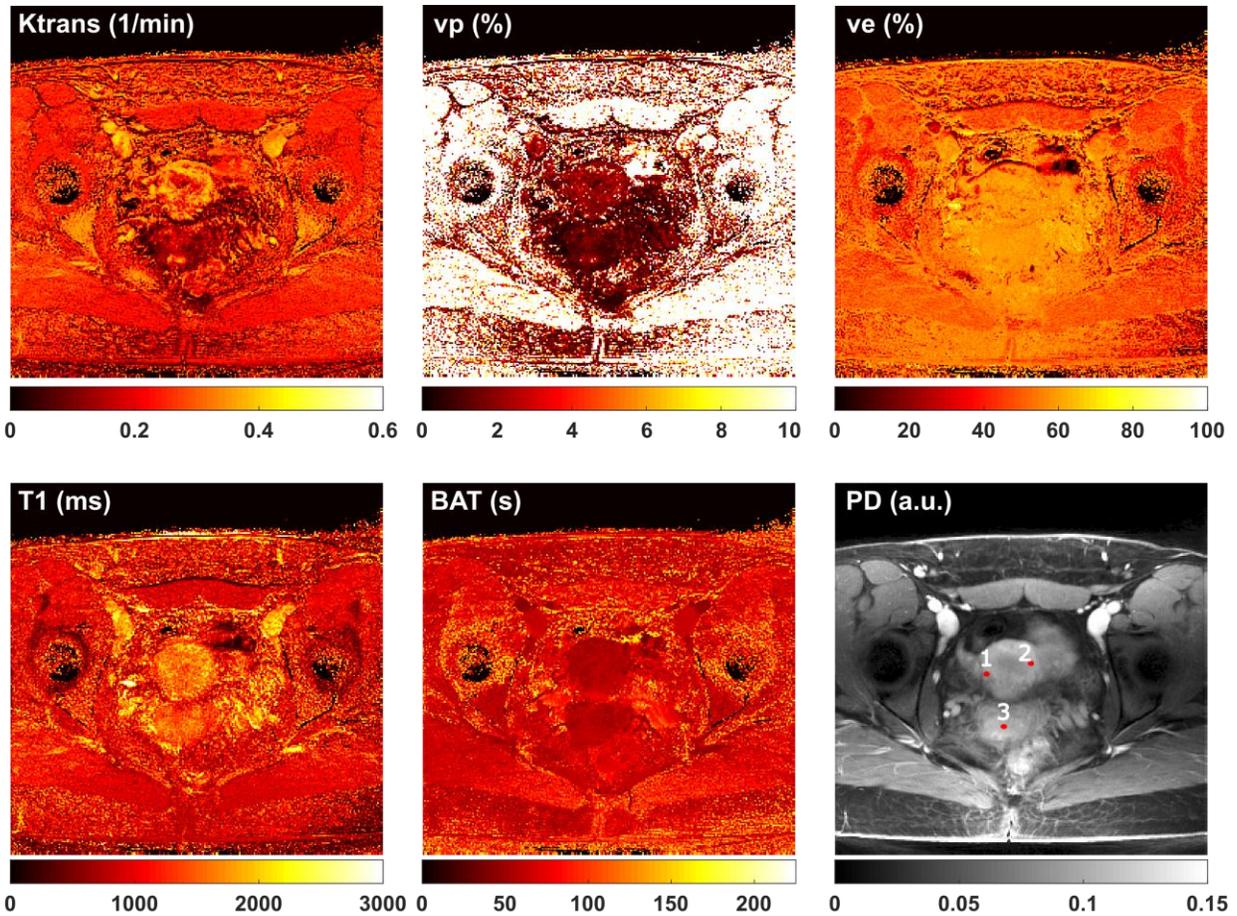

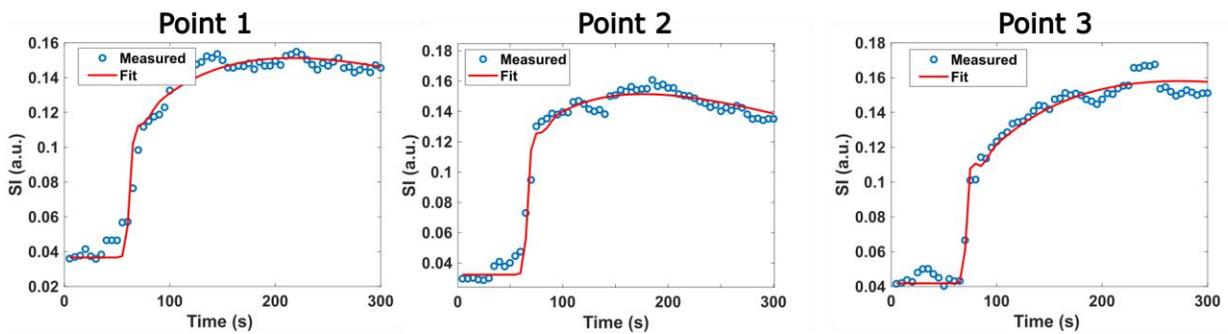

Figure 4: (A) Quantitative $K^{trans}$, $v_p$, $v_e$, T1, BAT and PD maps estimated by DCE-Qnet from a representative healthy subject. The three numbered red points on the PD map correspond to the sample concentration-time curves displayed in (B). (B) A comparison between the measured concentration-time curves (blue dots) and curves synthesized from the quantitative maps estimated by DCE-Qnet. Note the close agreement between the measured data and the synthesized curves.

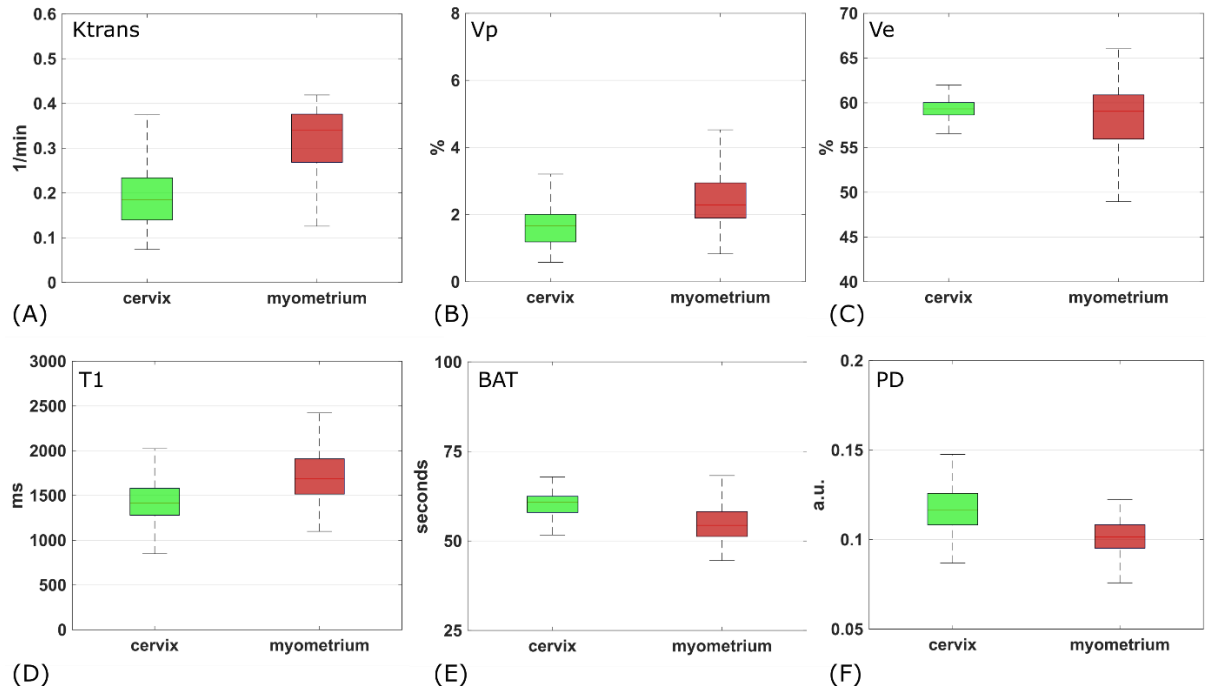

Figure 5: Box and whisker plots of the distribution of parameter values estimated with the proposed DCE-Qnet for different tissue types. Shown are (A) Ktrans, (B) $v_p$, (C) $v_e$, (D) T1, (E) BAT and (F) PD values.

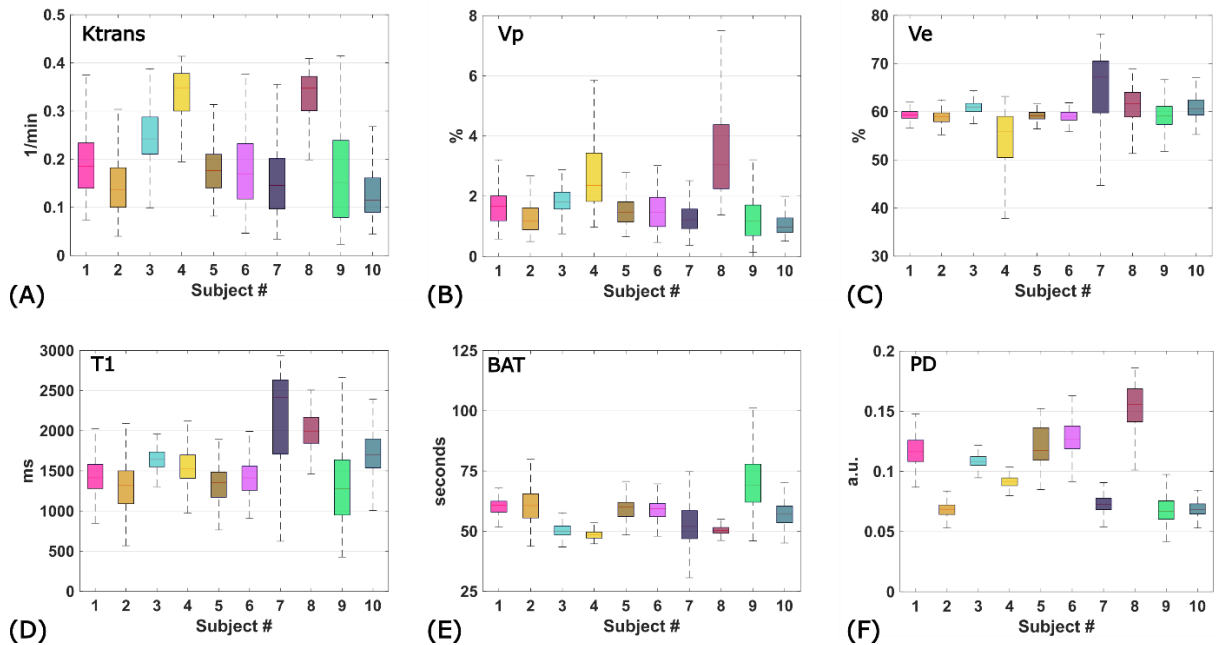

Figure 6: DCE and tissue parameter values for all healthy subjects scanned. The data was calculated for the ROI located in the cervix. Shown are (A) Ktrans, (B) $v_p$, (C) $v_e$, (D) T1, (E) BAT and (F) PD values.

### 3.2.2. Tumor patient

The quantitative parameter maps for each time point are shown in Figure 7A and the concentration-time curves of three representative tumor voxels in Figure 7B. The measured and synthesized curves were in strong agreement (r ≥ 0.99). There were notable differences in the maps of the two time points (Figure 7A) due to distinct patient positioning and bladder filling. The distribution of values of each estimated parameter within the tumor is shown in Figure 8. The mean±SD and CV for each parameter in the myometrium and the tumor are listed in Table 1. The test-retest repeatability varied by parameter yielding a CV in the tumor that ranged between 0.4 and 48%. A comparison of the results obtained in this study and those reported in the literature is listed in Table 2.

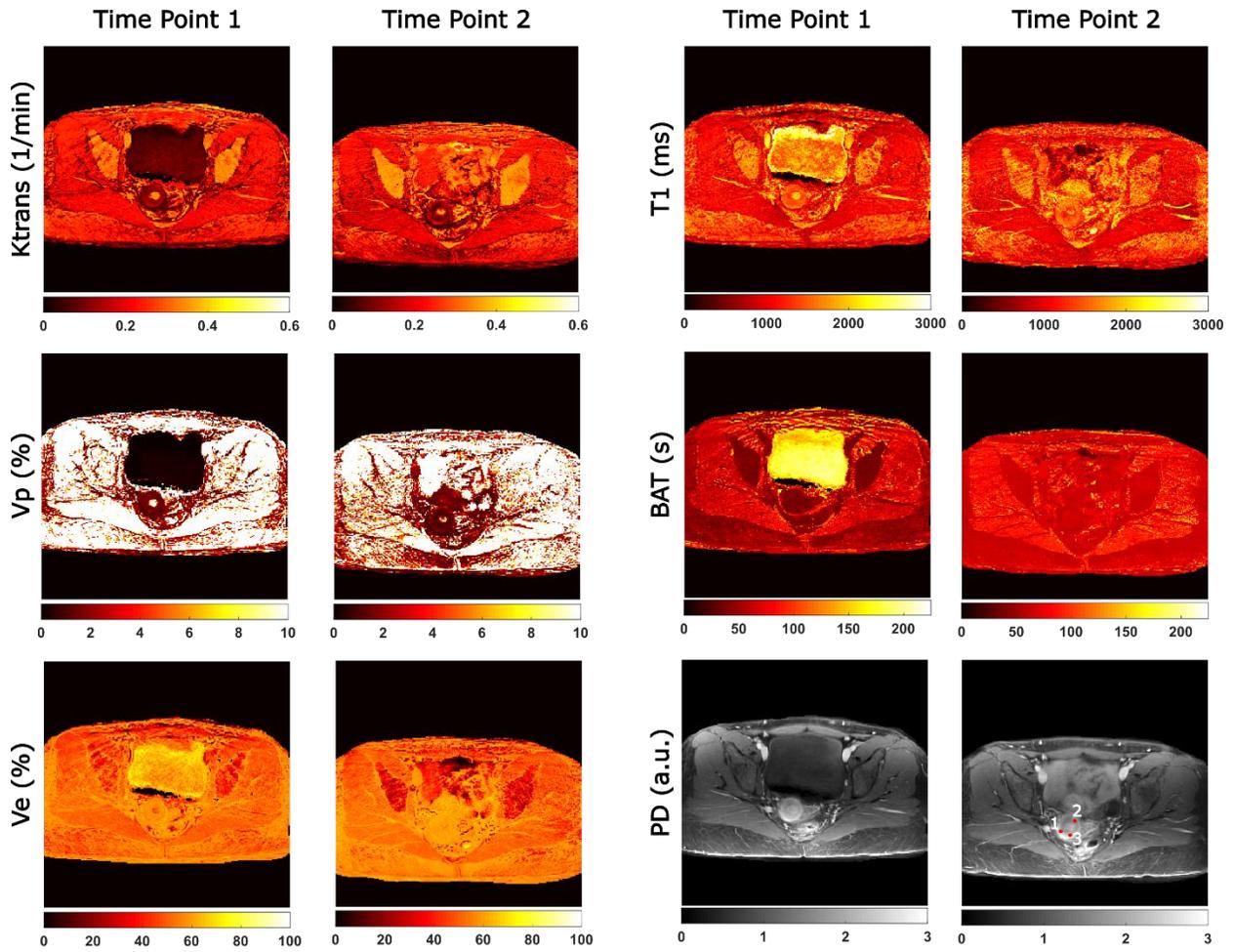
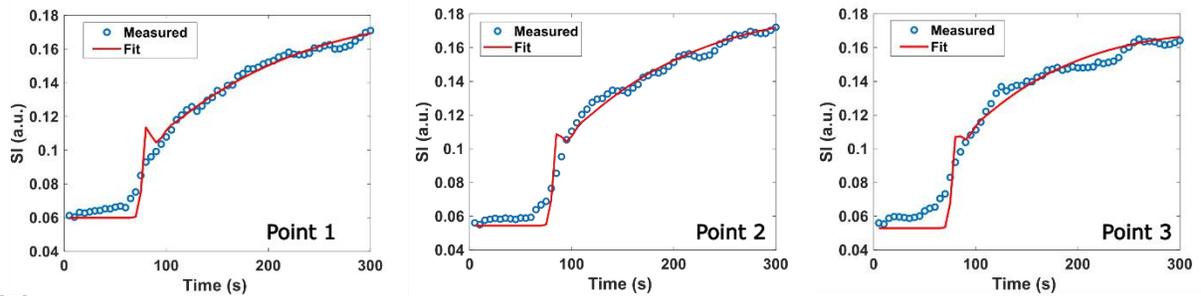

Figure 7: (A) Estimated parameter maps for the patient with tumor for the two time points scanned with the second time point occurring 24 hours after the initial scan. The three numbered red points on the PD map of the second time-point correspond to the sample concentration-time curves displayed in (B). (B) A comparison between the measured concentration-time curves (blue dots) and curves synthesized from the quantitative maps estimated by DCE-Qnet in the tumor ROI.

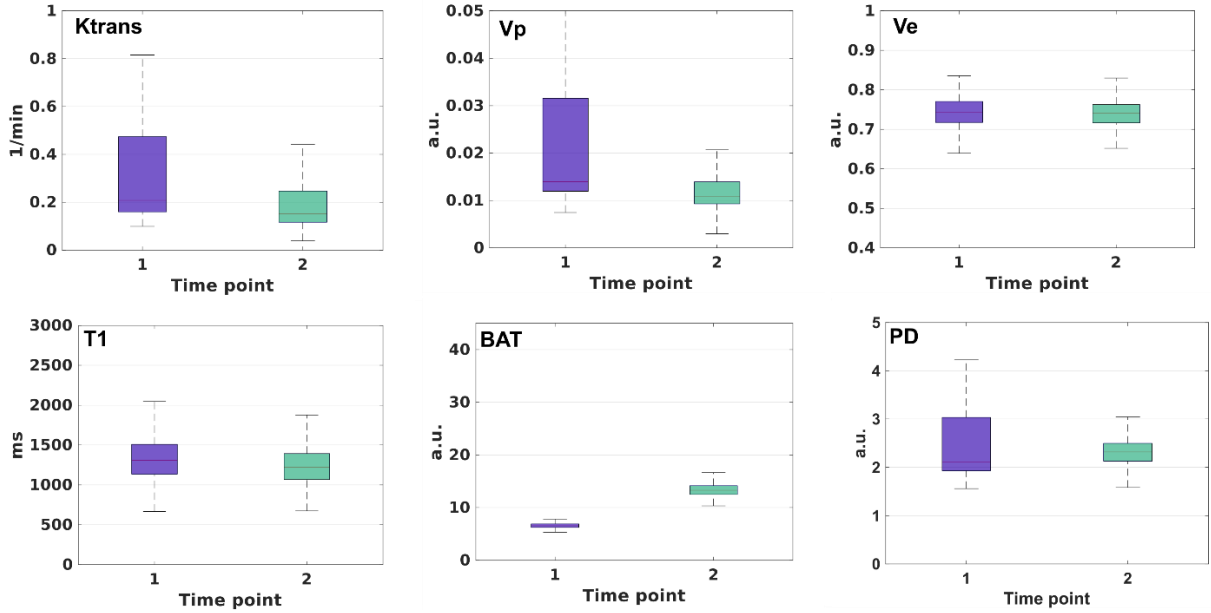

Figure 8: Box and whisker plots for the parameter values estimated using the proposed DNN in the tumor ROI for each of the measured time points. The second time point was acquired 24 hours after the first time point.

*Table 1: Mean±SD of tissue parameter values in the subjects scanned*

| Anatomical region | Tissue Parameters | | | | | |
|---|---|---|---|---|---|---|
| | $K^{trans}$ [1/min] | $V_p$ [%] | $V_e$ [%] | $T_1$ [ms] | BAT [sec] | PD [a.u.] |
| [a]*Healthy subjects* | | | | | | |
| *Cervix* | 0.21±0.07 | 2±1 | 59±3 | 1596±293 | 55±5 | 0.1±0.03 |
| *CV [%]* | 35.7 | 51.4 | 5.3 | 18.3 | 11.8 | 30.6 |
| *Uterine myometrium* | 0.28±0.06 | 4±3 | 56±5 | 1624±159 | 55±5 | 0.11±0.02 |
| *CV [%]* | 23.1 | 88.4 | 8.9 | 9.8 | 8.1 | 23.5 |
| [b]*Tumor patient* | | | | | | |
| *Uterine myometrium* | 0.25±0.03 | 8±5 | 6±1 | 1706±219 | 50±25 | 1.43±0.27 |
| *CV [%]* | 14.4 | 60.2 | 1.9 | 12.8 | 46.8 | 18.5 |
| *Tumor* | 0.18±0.02 | 4±2 | 58±3 | 1560±21 | 50±20 | 1.9±0.1 |
| *CV [%]* | 11.9 | 46.7 | 4.5 | 1.4 | 37.8 | 5.5 |

[a] CV calculated for the 10 subjects scanned

[b] CV calculated for the 2 time points scanned

# 4. Discussion

## 4.1. Numerical simulations

### 4.1.1. Effect of injected noise in training

The amount of noise injected in training had a measurable effect on the accuracy and image quality of the quantitative maps. By randomly varying the injected noise in each training epoch, the network learns to associate tissue parameters with signal features that exceed the noise floor rather than noise or artifacts (due to undersampling, for example). As such, increases in SNR did not improve the accuracy past a certain threshold (Figure 3). Increasing the injected noise reduced this threshold but also increased the absolute error. This can be understood to result from the noise overwhelming small signals in the phantom. Of note, the data in Figure 3 was generated with noise SD defined as a percentage of the *maximum* signal in the training dataset. This allows a 'worse-case' analysis but understandably results in relatively large errors, particularly for the case of 1 dB training SNR (Figure 3C). The error in vivo (Figure 4, Figure 7, Figure S2) was significantly smaller since a fixed noise SD is used instead.

For a given noise level, the error as a function of SNR differed between the quantified tissue parameters. For example, the error in the BAT was consistently lower than that of the other parameters. This is reflective of the inherent sensitivity of the sequence to different parameters which is a challenge for accurate DCE quantification as discussed below.

### 4.1.2. Comparison to conventional quantification

DCE quantification is commonly accomplished by NLSQ fitting of the measured data [23]. However, these fitting algorithms are known to be sensitive to initialization and noise and are more susceptible to getting trapped in local minima which affects the accuracy of the results (Figures 2B, 2C, Supporting Figure S2). DCE-Qnet quantification outperformed NLSQ fitting despite simultaneous estimation of a larger set of parameters (Figure 2A, Figure 2B). DCE-Qnet was trained on data generated by sampling the entire parameter space, which removes the need for parameter initialization and makes the quantification less vulnerable to local minima. Because it was trained on noisy data, DCE-Qnet was more robust to noisy measurements in vivo, enabling better delineation of the anatomy than NLSQ fitting even when separate B1 and T1 maps were provided to the NLSQ algorithm (Supporting Figure S2).

### 4.2. In vivo experiments

#### 4.2.1. Inter-subject variability

The quantified tissue parameters values exhibited significant variations across subjects (Figure 6, Table 1) in the cervix and myometrium. This is to be expected given the substantial influence of age and menstrual period on the MRI characteristics of the uterus. Moreover, subject position (e.g., anterversion, mid-positioned, vs. retroversion) and the relationship with adjacent pelvic organs such as the degree of distension of the bladder can also affect the resulting maps [24], [25], [26].

#### 4.2.2. In vivo reproducibility in the tumor

The in vivo reproducibility in the tumor varied by parameter. Most parameters ($K^{trans}$, $v_e$, T1 and PD) showed good to excellent reproducibility. Since the BAT depends on the bolus injection time, which is operator dependent, larger variations (i.e. poor reproducibility) are to be expected. The reproducibility of the $v_p$ was also limited. This may be due to several factors. First, the magnitude of the $v_p$ in the tumor was small and hence more likely affected by estimation noise. Second, although the test-retest experiments yielded maps that were visually similar, there were clear differences between the two scans due to patient positioning and bladder filling (Figure 7A). Bladder filling can deform neighboring organs thereby affecting the reproducibility of the measurements. Future studies will ensure consistent bladder filling between time points and will include more patients and more time points to guarantee accurate measurement of the reproducibility.

#### 4.2.3. Comparison with previous studies

Few studies have reported quantitative DCE parameters in healthy subjects due to growing awareness of the potential adverse effects of contrast agent accumulation [27]. However, the T1 values measured in this study are similar to those reported by Bazelaire et al [28] in the cervix (1596±293 vs 1616±61 ms) and myometrium (1624±159 vs 1514±146 ms) which serves to validate our approach.

There is a wide range of reported $K^{trans}$, $v_p$ and $v_e$ values in cervical tumors, as shown in Table 2. Although some of these differences may be due to tumor heterogeneity, it is certain that differences in experimental conditions contributed to the differences in the values as well. Differences in the selection of AIF, hematocrit or relaxivity values assumed and the BAT processing used can affect the quantification. To our knowledge, this study is the first to account for the effects of B1 inhomogeneity in the DCE quantification. This is important because T1 maps are commonly obtained using a variable flip angle (VFA) sequence which is strongly affected by $B_1$ inhomogeneity [29], [30]. In turn, this affects the concentration-time curves leading to inaccuracies in the quantification. The use of DNN for DCE quantification has been proposed by others [31] though there are notable differences to our approach. First, this work proposes a joint estimation of the perfusion parameters along with T1 and B1. Secondly, DCE-Qnet is trained entirely on simulated data which eliminates the need for patient data collection. Additionally, the patient data used in this study was acquired with a relatively slow (5 seconds/frame) sampling rate, leading to a more challenging quantification. DCE-Qnet was nevertheless able to successfully reconstruct the quantitative tissue maps. Finally, unlike Ref. [31], the clinical focus of this study is cervical cancer.

### 4.3. Pharmacokinetic model

This study used the Extended-Tofts pharmacokinetic model which can be motivated by consideration of the relatively slow sampling rate in our acquisition in comparison to e.g. Donaldson et al [32] and the assumption of highly perfused cervical tissue [33]. The use of the Tofts model also enables comparison with the numerous other studies listed in Table 2 that use this model. While various DCE models exist in the literature [19], the goal of this study was not to evaluate the merits of different models but to demonstrate the feasibility of DCE quantification using a neural network trained exclusively with simulated data. Indeed, DCE-Qnet is model-agnostic and can be applied with other pharmacokinetic models by simply changing the training dataset used.

### 4.4. T1 estimation from contrast enhancement

DCE-Qnet introduces a novel way to estimate T1 from contrast enhancement only, which does not need variation of the flip angle in the acquisition as in standard T1 estimation techniques. Instead, the network is trained to exploit the dynamic T1 reduction produced by the passage of the contrast agent to infer the (contrast-free) tissue T1. This approach eliminates the need for a separate T1 scan and ensures perfect registration between all mapped parameters. However, since the T1 sensitivity is dependent on tracer uptake, T1 values in regions with poor uptake may incur errors in the estimation.

### 4.5. Challenges of deep learning for DCE quantification

Deep learning quantification using physics-derived simulated training data was originally proposed for T1 and T2 estimation using MRF [6] and later adapted to quantification of other highly-multiparametric acquisitions [7], [8], [34], [35]. The key advantages of DNN-based quantification include rapid computation, elimination of the need for patient data, and simultaneous estimation of multiple parameters. Such features are beneficial in DCE quantification as well but limitations inherent to the DCE acquisition require careful consideration. In MRF, scan parameters can be freely varied to induce a differential signal evolution between tissues with different underlying parameters. In contrast, the DCE-MRI signal is defined by the contrast agent uptake in the tissue which is physiologically driven and only weakly dependent on the scan parameters. This limits the achievable sensitivity of the acquisition to small variations in tissue parameters. Tissues with low contrast agent uptake, such as fat, will therefore have a small DCE signal potentially resulting in larger errors in the estimated parameters for that type of tissues.

### 4.6. Limitations and future work

This study used the population-based Parker AIF [15], [36] to improve reproducibility and to simplify the comparison to previous studies [32], [37], [38], [39], [40], [41]. Nevertheless, a subject-specific AIF may yield more accurate DCE parameters. Existing subject-specific AIF measurement methods require manual intervention and can suffer from partial volume and motion effects. Although outside the scope of this study, the proposed framework can also be adapted to include an estimation of subject-specific AIF along with the other tissue parameters. This is left for future studies. It should also be noted that T1 and B1 are covariates so errors in the estimation of one will propagate to the other parameters, reducing the precision. However, this can be mitigated by a separate B1 acquisition.

Table 2: Cervical tumor DCE parameter values reported in the literature

| Reference | No. of subjects | Field (T) | Contrast Agent | Model | AIF | $K^{trans}$ | $V_p$ [%] | $V_e$ [%] |
|---|---|---|---|---|---|---|---|---|
| **This study** | 10 (healthy) 1 (tumor) | 3 | Magnevist | Extended Tofts | Parker | 0.21±0.07 (healthy) 0.18±0.02 (tumor) | 2±1 (healthy) 4±2 (tumor) | 59±3 (healthy) 58±3 (tumor) |
| [a]Kallehauge et al., 2013 [37] | 16 | 3 | Dotarem | Extended Tofts | Various | 0.46±0.29 | 6±5 | 54±14 |
| [b]Liu et al., 2020, [38] | 48 | 3 | Omniscan | Extended Tofts | Image-derived | 1.24±0.34 | 64±21 | 14±9 |
| Andersen et al., 2011, [39] | 78 | 1.5 | Magnevist | Tofts | Population based | 0.18±0.06 | ---- | 42±16 |
| Donaldson et al., 2010, [32] | 30 | 1.5 | Magnevist | Extended Tofts | Image-derived | 0.35±0.26 | 4±3 | 34±17 |
| Yang et al., 2010, [40] | 39 | 1.5 | Omniscan | Extended Tofts | Parker | 0.18±0.05 | 2±1 | 36±11 |
| [c]Feng et al., 2020, [41] | 38 | 1.5 | Gado-diamide | Extended Tofts | Image-derived | 0.71±0.32 | 73±15 | 24±21 |

[a] average of Parker, magnitude and CT-derived AIFs, [b] average of non-residual and residual group, [c] average of significant and non-significant reaction groups

## 5. Conclusion

This work describes a novel deep learning approach for comprehensive DCE-MRI quantification from a single acquisition. The method accounts for B1 inhomogeneity and simultaneously computes 6 tissue parameter maps, eliminating the need for separate T1 acquisition or BAT processing leading to a 10 minutes reduction in scan time and more accurate and reproducible quantification.

## 6. Appendix

Contrast agent uptake was represented using the Extended Tofts model (ETM) [36], [42] where changes in concentration are modeled as a function of time. The AIF is the concentration of the contrast agent in the plasma of a feeding artery and is denoted by $C_b(t)$. The AIF was obtained from the population-derived Parker AIF model [15]. The blood plasma concentration of the contrast agent, $C_p(t)$, was obtained by scaling the AIF by the hematocrit (Hct) concentration: $C_p(t) = C_b(t)/(1-Hct)$, with Hct=0.42.

Defining $v_p$ as the plasma volume fraction, $v_e$ as the extracellular extravascular space (EES) in tissue and $K^{trans}$ as the volume transfer constant from plasma to EES, the tissue concentration of CA is given by [36], [42]:

$$C_t(t) = v_p C_p(t) + K^{trans} \int_0^t C_p(t') \exp\left(\frac{-K^{trans}(t-t')}{v_e}\right) dt' \qquad (A1)$$

The arrival delay of the CA bolus to the tissue is modeled as a shift of the $C_t(t)$ curve by the BAT:

$$C_t^s(t) = C_t(t - BAT) \qquad (A2)$$

The presence of CA induces a reduction in the $T_1$ from its native value ($T_{10}$) that is proportionate to the concentration-time curve and the relaxivity ($r_1$) of the CA:

$$\frac{1}{T_1} = \frac{1}{T_{10}} + r_1 C_t^s(t) \qquad (A3)$$

The T1 reduction affects the measured MRI signal, which for a spoiled gradient echo sequence, is dependent on several additional parameters such as the equilibrium signal $S_0$, the TR and the FA α as well as the transmitter magnetic field inhomogeneity $B_1$:

$$S = S_0 \sin(B_1 \cdot \alpha) \frac{1-\exp\left(-\frac{TR}{T_1}\right)}{1-\cos(B_1\cdot\alpha)\cdot\left(\exp\left(-\frac{TR}{T_1}\right)\right)} \tag{A4}$$

The concentration-dependent signal equation can be obtained by substituting Eqn. (3) into Eqn. (A4):

$$S = S_0 \sin(B_1 \cdot \alpha) \frac{1-\exp\left(-TR\left(\frac{1}{T_{10}}+r_1 C_t^S(t)\right)\right)}{1-\cos(B_1\cdot\alpha)\cdot\exp\left(-TR\left(\frac{1}{T_{10}}+r_1 C_t^S(t)\right)\right)} \tag{A5}$$

During training, the DNN learns a mapping between the DCE parameters, T1, BAT, and the measured signal as defined by Eq. (A5).

## 7. Acknowledgements

Research reported in this publication was supported by National Cancer Institute / National Institutes of Health awards R01-CA244532, R37-CA262662 and P30-CA008748. The content is solely the responsibility of the authors and does not necessarily represent the official views of the National Institutes of Health.

## 8. Authors' Contributions

Ouri Cohen: (Study conception and design, Analysis and interpretation of data, Drafting of manuscript, Critical revision), Soudabeh Kargar: (Study conception and design, Acquisition of data, Analysis and interpretation of data), Ricardo Otazo: (Study conception and design, Analysis and interpretation of data, Drafting of manuscript, Critical revision), Alberto Vargas: (Study conception and design, Analysis and interpretation of data), Sungmin Woo: (Study conception and design, Analysis and interpretation of data, Critical revision)

## 9. Declaration of competing interest

Ricardo Otazo has a patent on the GRASP technique.

## 10. Data availability statement

The data that support the findings of this study are not openly available due to patient privacy requirements.

## 11. References

[1] A. R. Padhani and K. A. Miles, "Multiparametric imaging of tumor response to therapy," *Radiology*, vol. 256, no. 2, pp. 348–364, 2010.
[2] J. P. O'Connor et al., "Imaging biomarker roadmap for cancer studies," *Nature reviews Clinical oncology*, vol. 14, no. 3, pp. 169–186, 2017.
[3] S. M. Galbraith et al., "Reproducibility of dynamic contrast-enhanced MRI in human muscle and tumours: comparison of quantitative and semi-quantitative analysis," *NMR in Biomedicine: An International Journal Devoted to the Development and Application of Magnetic Resonance In Vivo*, vol. 15, no. 2, pp. 132–142, 2002.
[4] Y. Bliesener, J. Acharya, and K. S. Nayak, "Efficient DCE-MRI parameter and uncertainty estimation using a neural network," *IEEE transactions on medical imaging*, vol. 39, no. 5, pp. 1712–1723, 2019.
[5] K. Hornik, "Some new results on neural network approximation," *Neural networks*, vol. 6, no. 8, pp. 1069–1072, 1993.
[6] O. Cohen, B. Zhu, and M. S. Rosen, "MR fingerprinting deep reconstruction network (DRONE)," *Magnetic resonance in medicine*, vol. 80, no. 3, pp. 885–894, 2018.
[7] O. Cohen et al., "CEST MR Fingerprinting (CEST-MRF) for Brain Tumor Quantification Using EPI Readout and Deep Learning Reconstruction," *Magnetic Resonance in Medicine*, 2022.
[8] O. Cohen and R. Otazo, "Global Deep Learning Optimization of CEST MR Fingerprinting (CEST-MRF) Acquisition Schedule," *NMR in Biomedicine*, p. e4954, 2023.
[9] G. Ongie, A. Jalal, C. A. Metzler, R. G. Baraniuk, A. G. Dimakis, and R. Willett, "Deep learning techniques for inverse problems in imaging," *IEEE Journal on Selected Areas in Information Theory*, vol. 1, no. 1, pp. 39–56, 2020.
[10] M. Genzel, J. Macdonald, and M. März, "Solving inverse problems with deep neural networks–robustness included?," *IEEE Transactions on Pattern Analysis and Machine Intelligence*, vol. 45, no. 1, pp. 1119–1134, 2022.
[11] A. Rastogi, A. Dutta, and P. K. Yalavarthy, "VTDCE-Net: A time invariant deep neural network for direct estimation of pharmacokinetic parameters from undersampled DCE MRI data," *Medical Physics*, vol. 50, no. 3, pp. 1560–1572, 2023.
[12] J. Zou, J. M. Balter, and Y. Cao, "Estimation of pharmacokinetic parameters from DCE-MRI by extracting long and short time-dependent features using an LSTM network," *Medical physics*, vol. 47, no. 8, pp. 3447–3457, 2020.
[13] J. Bae, C. Li, A. Masurkar, Y. Ge, and S. G. Kim, "Improving measurement of blood-brain barrier permeability with reduced scan time using deep-learning-derived capillary input function," *NeuroImage*, vol. 278, p. 120284, 2023.
[14] J.-S. Park, "Optimal Latin-hypercube designs for computer experiments," *Journal of statistical planning and inference*, vol. 39, no. 1, pp. 95–111, 1994.
[15] G. J. Parker et al., "Experimentally-derived functional form for a population-averaged high-temporal-resolution arterial input function for dynamic contrast-enhanced MRI," *Magnetic Resonance in Medicine: An Official Journal of the International Society for Magnetic Resonance in Medicine*, vol. 56, no. 5, pp. 993–1000, 2006.
[16] D. P. Kingma and J. Ba, "Adam: A method for stochastic optimization," *arXiv preprint arXiv:1412.6980*, 2014.
[17] R. M. Zur, Y. Jiang, L. L. Pesce, and K. Drukker, "Noise injection for training artificial neural networks: A comparison with weight decay and early stopping," *Medical physics*, vol. 36, no. 10, pp. 4810–4818, 2009.

[18] S. Metzner, G. Wübbeler, S. Flassbeck, C. Gatefait, C. Kolbitsch, and C. Elster, "Bayesian uncertainty quantification for magnetic resonance fingerprinting," *Physics in Medicine & Biology*, vol. 66, no. 7, p. 075006, 2021.

[19] S. P. Sourbron and D. L. Buckley, "Classic models for dynamic contrast-enhanced MRI," *NMR in Biomedicine*, vol. 26, no. 8, pp. 1004–1027, 2013.

[20] L. Feng *et al.*, "Golden-angle radial sparse parallel MRI: combination of compressed sensing, parallel imaging, and golden-angle radial sampling for fast and flexible dynamic volumetric MRI," *Magnetic resonance in medicine*, vol. 72, no. 3, pp. 707–717, 2014.

[21] L. Feng, L. Axel, H. Chandarana, K. T. Block, D. K. Sodickson, and R. Otazo, "XD-GRASP: golden-angle radial MRI with reconstruction of extra motion-state dimensions using compressed sensing," *Magnetic resonance in medicine*, vol. 75, no. 2, pp. 775–788, 2016.

[22] L. I. Sacolick, F. Wiesinger, I. Hancu, and M. W. Vogel, "B1 mapping by Bloch-Siegert shift," *Magnetic resonance in medicine*, vol. 63, no. 5, pp. 1315–1322, 2010.

[23] T. S. Ahearn, R. T. Staff, T. W. Redpath, and S. I. K. Semple, "The use of the Levenberg–Marquardt curve-fitting algorithm in pharmacokinetic modelling of DCE-MRI data," *Physics in Medicine & Biology*, vol. 50, no. 9, p. N85, 2005.

[24] D. Well *et al.*, "Age-related structural and metabolic changes in the pelvic reproductive end organs," in *Seminars in nuclear medicine*, Elsevier, 2007, pp. 173–184.

[25] K. Angelopoulos *et al.*, "Computed tomography contrast enhancement pattern of the uterus in premenopausal women in relation to menstrual cycle and hormonal contraception," *Acta Radiologica*, vol. 62, no. 9, pp. 1257–1262, 2021.

[26] J. E. Langer, E. R. Oliver, A. S. Lev-Toaff, and B. G. Coleman, "Imaging of the female pelvis through the life cycle," *Radiographics*, vol. 32, no. 6, pp. 1575–1597, 2012.

[27] J. Ramalho, R. Semelka, M. Ramalho, R. Nunes, M. AlObaidy, and M. Castillo, "Gadolinium-based contrast agent accumulation and toxicity: an update," *American Journal of Neuroradiology*, vol. 37, no. 7, pp. 1192–1198, 2016.

[28] C. M. De Bazelaire, G. D. Duhamel, N. M. Rofsky, and D. C. Alsop, "MR imaging relaxation times of abdominal and pelvic tissues measured in vivo at 3.0 T: preliminary results," *Radiology*, vol. 230, no. 3, pp. 652–659, 2004.

[29] M. C. Schabel and D. L. Parker, "Uncertainty and bias in contrast concentration measurements using spoiled gradient echo pulse sequences," *Physics in Medicine & Biology*, vol. 53, no. 9, p. 2345, 2008.

[30] C. Roberts, R. Little, Y. Watson, S. Zhao, D. L. Buckley, and G. J. Parker, "The effect of blood inflow and B1-field inhomogeneity on measurement of the arterial input function in axial 3D spoiled gradient echo dynamic contrast-enhanced MRI," *Magnetic resonance in medicine*, vol. 65, no. 1, pp. 108–119, 2011.

[31] T. Ottens *et al.*, "Deep learning DCE-MRI parameter estimation: Application in pancreatic cancer," *Medical Image Analysis*, vol. 80, p. 102512, 2022.

[32] S. B. Donaldson *et al.*, "A comparison of tracer kinetic models for T1-weighted dynamic contrast-enhanced MRI: Application in carcinoma of the cervix," *Magnetic Resonance in Medicine*, vol. 63, no. 3, pp. 691–700, 2010.

[33] S. P. Sourbron and D. L. Buckley, "On the scope and interpretation of the Tofts models for DCE-MRI," *Magnetic resonance in medicine*, vol. 66, no. 3, pp. 735–745, 2011.

[34] Q. Zhang *et al.*, "Deep learning–based MR fingerprinting ASL ReconStruction (DeepMARS)," *Magnetic resonance in medicine*, vol. 84, no. 2, pp. 1024–1034, 2020.

[35] O. Perlman *et al.*, "Quantitative imaging of apoptosis following oncolytic virotherapy by magnetic resonance fingerprinting aided by deep learning," *Nat. Biomed. Eng.*, 2021, doi: 10.1038/s41551-021-00809-7.
[36] G. J. Parker and D. L. Buckley, "Tracer kinetic modelling for T 1-weighted DCE-MRI," *Dynamic contrast-enhanced magnetic resonance imaging in oncology*, pp. 81–92, 2005.
[37] J. Kallehauge *et al.*, "Voxelwise comparison of perfusion parameters estimated using dynamic contrast enhanced (DCE) computed tomography and DCE-magnetic resonance imaging in locally advanced cervical cancer," *Acta Oncologica*, vol. 52, no. 7, pp. 1360–1368, 2013.
[38] B. Liu *et al.*, "DCE-MRI quantitative parameters as predictors of treatment response in patients with locally advanced cervical squamous cell carcinoma underwent CCRT," *Frontiers in Oncology*, vol. 10, p. 585738, 2020.
[39] E. K. Andersen, G. B. Kristensen, H. Lyng, and E. Malinen, "Pharmacokinetic analysis and k-means clustering of DCEMR images for radiotherapy outcome prediction of advanced cervical cancers," *Acta Oncologica*, vol. 50, no. 6, pp. 859–865, 2011.
[40] C. Yang, W. M. Stadler, G. S. Karczmar, M. Milosevic, I. Yeung, and M. A. Haider, "Comparison of quantitative parameters in cervix cancer measured by dynamic contrast–enhanced MRI and CT," *Magnetic Resonance in Medicine: An Official Journal of the International Society for Magnetic Resonance in Medicine*, vol. 63, no. 6, pp. 1601–1609, 2010.
[41] Y. Feng *et al.*, "Combined dynamic DCE-MRI and diffusion-weighted imaging to evaluate the effect of neoadjuvant chemotherapy in cervical cancer," *Tumori Journal*, vol. 106, no. 2, pp. 155–164, 2020.
[42] P. S. Tofts, "Modeling tracer kinetics in dynamic Gd-DTPA MR imaging," *Journal of magnetic resonance imaging*, vol. 7, no. 1, pp. 91–101, 1997.

## 12. Figure Captions

**Figure 1:** Main components of the DCE-Qnet estimation process. (A) Sampling of the DCE parameters shown for 3 out of 6 parameters for visualization purposes. The parameter values were generated by LH sampling of the 6-dimensional space. (B) The population-based Parker AIF was used to generate the concentration-time curves for each of the DCE parameter combinations. (C) Concentration-time curves were converted to signal intensity values using the signal model defined in Eqn. (5). (D) Diagram of the neural network used. Signal intensity curves generated from the sampled DCE parameters were used as input to the network during training. The network learned to map the signal intensity curves to the underlying tissue parameters. The proton density was then calculated from measured and synthesized signals obtained with the estimated DCE parameters.

**Figure 2:** Reconstruction accuracy and comparison between NLSQ and DCE-Qnet. (A) Estimated vs reference comparison for parameters estimated with DCE-Qnet for a known fixed B1 equivalent to using a measured B1 map. (B) Estimated vs reference comparison for simultaneous DCE-Qnet estimation of all parameters. Although the variance was larger, the DCE-Qnet values were nevertheless strongly correlated with the reference values despite simultaneous estimation of all 6 parameters. (C) Estimated vs reference comparison for joint estimation of all parameters using the NLSQ algorithm. Compare with the DCE-Qnet estimation shown in (B). (D) NLSQ estimation for fixed B1. (E) NLSQ estimation when the true B1, true T1 and true BAT were provided as inputs to the function.

**Figure 3:** Normalized RMS error (NMRE) for different parameters as a function of SNR for networks trained with different injected noise levels. Shown is the output from the network trained with (A) 0.1% noise, (B) 1% noise and (C) 10%. Note that higher levels of injected noise reduced the SNR at which the error converged albeit at the cost of a larger error.

**Figure 4:** (A) Quantitative Ktrans, $v_p$, $v_e$, T1, BAT and PD maps estimated by DCE-Qnet from a representative healthy subject. The three numbered red points on the PD map correspond to the sample concentration-time curves displayed in (B). (B) A comparison between the measured concentration-time curves (blue dots) and curves synthesized from the quantitative maps estimated by DCE-Qnet. Note the close agreement between the measured data and the synthesized curves.

**Figure 5:** Box and whisker plots of the distribution of parameter values estimated with the proposed DCE-Qnet for different tissue types. Shown are (A) Ktrans, (B) $v_p$, (C) $v_e$, (D) T1, (E) BAT and (F) PD values.

**Figure 6:** DCE and tissue parameter values for all healthy subjects scanned. The data was calculated for the ROI located in the cervix. Shown are (A) Ktrans, (B) $v_p$, (C) $v_e$, (D) T1, (E) BAT and (F) PD values.

**Figure 7:** (A) Estimated parameter maps for the patient with tumor for the two time points scanned with the second time point occurring 24 hours after the initial scan. The three numbered red points on the PD map of the second time-point correspond to the sample concentration-time curves displayed in (B). (B) A comparison between the measured concentration-time curves

(blue dots) and curves synthesized from the quantitative maps estimated by DCE-Qnet in the tumor ROI.

**Figure 8:** Box and whisker plots for the parameter values estimated using the proposed DNN in the tumor ROI for each of the measured time points. The second time point was acquired 24 hours after the first time point.

**Supporting Information**

**Supporting Figure S1:** Estimated vs reference tissue parameter values for a network trained with the correct (blue points) and wrong (red points) acquisition parameters. The network trained with the wrong acquisition parameters resulted in increased error as indicated by the higher RMSE.

**Supporting Figure S2:** (A) Tissue maps obtained with NLSQ fitting using measured B1 and T1 maps. (B) Tissue maps obtained with DCE-Qnet for the same subject. DCE-Qnet yielded maps with improved image quality due to the greater robustness to noise imparted by the network training. (C) Distribution of tissue parameter values for each method in an ROI placed in the cervix.

**Supporting Table S3:** Statistical measures of the agreement between the true and estimated parameter values for the different fitting experiments shown in Figure 2.